**The Climate Cost of Climate Investment: A Two-Period Perspective**

*Rohan Dubey* and *Shaunak Kulkarni*

August 26, 2024

Working paper for public distribution.

## Introduction

A major barrier to international cooperation in the realm of international climate action is an apparent lack of fair and just allocation of responsibility between stakeholder groups. Understanding, isolating, and instituting mechanisms to manage incentives against sharing responsibility can go a long way in sustaining effective long-term cooperation. The campaign against climate change has not been all blue skies and fair winds, and differences are beginning to show. Patience runs thin in the face of calls for more deliberate action due to the lack of visible results; national tempers have begun to flare, and widening fault lines are beginning to hamper effective collaboration (Moos & Arndt, 2023).

The main aim of this study is to establish whether climate investment demonstrates diminishing marginal return in terms of carbon productivity, and to evaluate the implications for climate action policy. In the interest of brevity and accuracy (given the exclusive use of secondary data) to minimise subjectivity of analysis, this study will be limited to exploring a binary relationship between the magnitude of cumulative investment and the expected (or observed) return on each additional unit of investment. Efforts by developed economies are very well documented, and often tend to be most visible in terms of awareness and scale; our research seeks to leverage this fact by first understanding whether there is a statistically significant relationship between the relative development of an economy and accumulated impact of climate investment, and using said findings to frame a thesis in terms of the international opportunity cost of domestic climate investment as a function of relative economic development.

This research has been divided into three minor studies, each building towards a final conclusion regarding the marginal productivity of climate investment, categorised as follows:

1. Assessing the relationship between cumulative impact of domestic climate investment and the domestic state of economic development.
2. Exploring links between accumulated impact of climate investment and the effectiveness of additional investment.
3. Discussing policy implications surrounding the climate impact of additional investment in terms of the domestic state of economic development.

### Quick Reference

## 1A – Assessing the relationship between the cumulative impact of climate investment and the state of economic development: Key Factors & Methodology

For the purpose of this study, carbon productivity will stand in as a relative indicator of the cumulative impact of climate investment, as it is a well-defined and quantifiable metric for which standardised records can be found. In this paper, we use 'the amount of GDP produced per unit of carbon equivalents emitted' (Beinhocker, et al., 2008) as the definition of carbon productivity.

Having defined a quantitative proxy for the cumulative impact of climate investment, we now move on to economic development, for which we use Human Development Index (HDI) to stand in for qualitative development and the relative well-being of economic actors. It would be prudent to note that we also considered State capacity, 'the ability of governments to effectively implement their policies and achieve their goals' (Herre, et al., 2023) to stand in for financial and administrative development as a supporting factor.

In the real world, economies with high state capacity are likely to exert their influence to improve domestic standards of living; additionally, indicators used to capture a state's administrative capacity may overlap with factors that influence HDI (Hanson & Sigman, 2021). We have therefore chosen to use HDI as the primary indicator of development, in line with conventional practice.

We begin by defining each variable and sourcing raw data for a model as follows:
*Carbon Productivity* $= \beta_0 + \beta_1 HDI + \beta_2 SCI + \epsilon$

1. HDI in arbitrary units, as an *Independent Variable* (UNDP, 2022)
2. State Capacity Index (SCI) in arbitrary units, as an *Independent Variable* (Herre, et al., 2023)
3. $CO_2$ equivalent emissions in kilograms, as a *Dependent Variable* (Ritchie, et al., 2020)
4. GDP in US$ as a *Dependent Variable* (World Bank, 2023)
5. Carbon Productivity as a composite *Dependent Variable*, $\frac{\text{GDP}}{\text{GHG}}$ in US$ per kilogram.

Next, we filter the raw data to eliminate inconsistent values (e.g. some combination of HDI, SCI, GDP, etc. absent) and assign each entry a unique (arbitrary) identifier that communicates country name and year of record, using abbreviations sourced with the data; e.g. records for Afghanistan from 2014 will be linked to AFG2014.

Building on the distinction between developing and developed economies, we test each to be fitted through a specific regression model, also tested for functional form misspecification using Ramsey's RESET test (see Appendix II for full results).

## 1B – Assessing the relationship between the cumulative impact of climate investment and the state of economic development: Inferences

The fundamental argument in favour of a proposition that the cumulative impact of climate investment is linked to the state of economic development lies in the very nature of 'development' as an economic ideal; given the criteria we have applied (HDI and SCI), a developed economy can be defined as one that is allocatively efficient and makes economically optimal decisions. Whether economic optimality emerges from effective markets, strategically sound decisions made in the past, or some combination of geopolitical factors, it stands to reason that high HDI and SCI scores are achieved by considering the economic 'big picture' which ultimately figures into national climate action policy.

Further, it stands to reason that relative economic optimality implies greater carbon productivity, whether by means of climate-conscious investment or simply efficient use of resources; carbon productivity is directly linked to wastage in an economy, such that wasteful practices will necessitate higher emissions by requiring greater energy use in the production process, either at the time of output or in producing additional quantities of input.

This positive relationship of HDI, SCI, and Carbon Productivity is illustrated through *Chart 1.1.* (HDI score and Carbon Productivity) and *Chart 1.2.* (SCI score and Carbon Productivity).

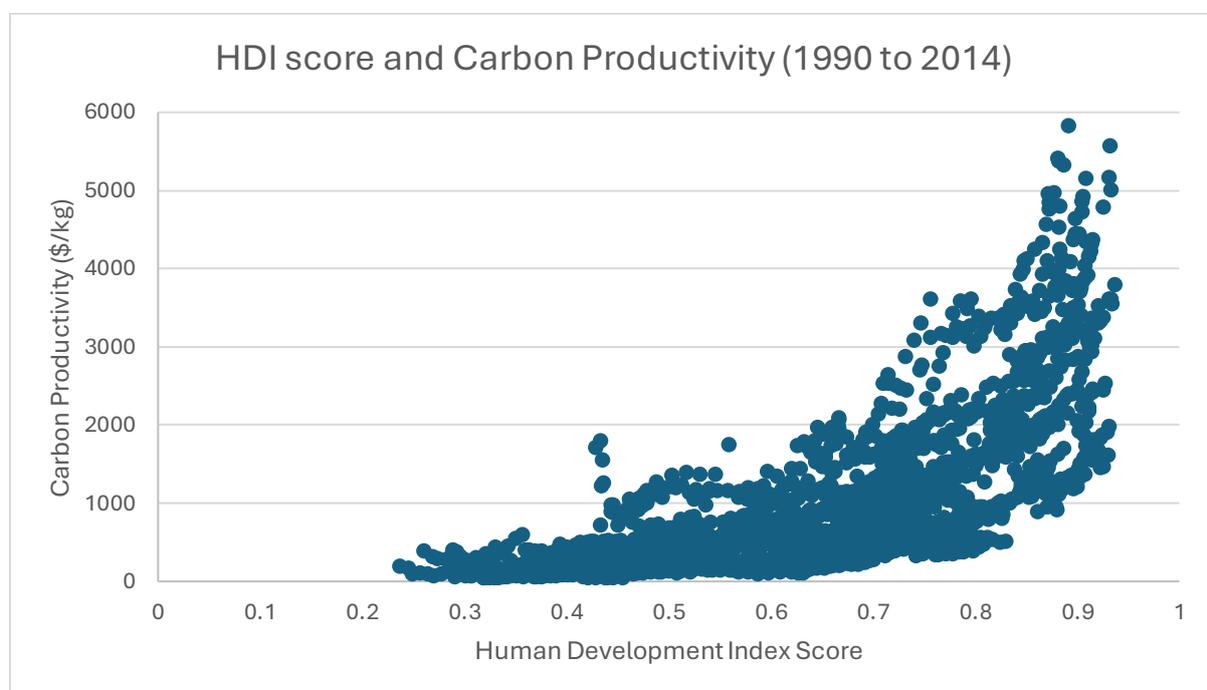

*Chart 1.1 : HDI score and Carbon Productivity*

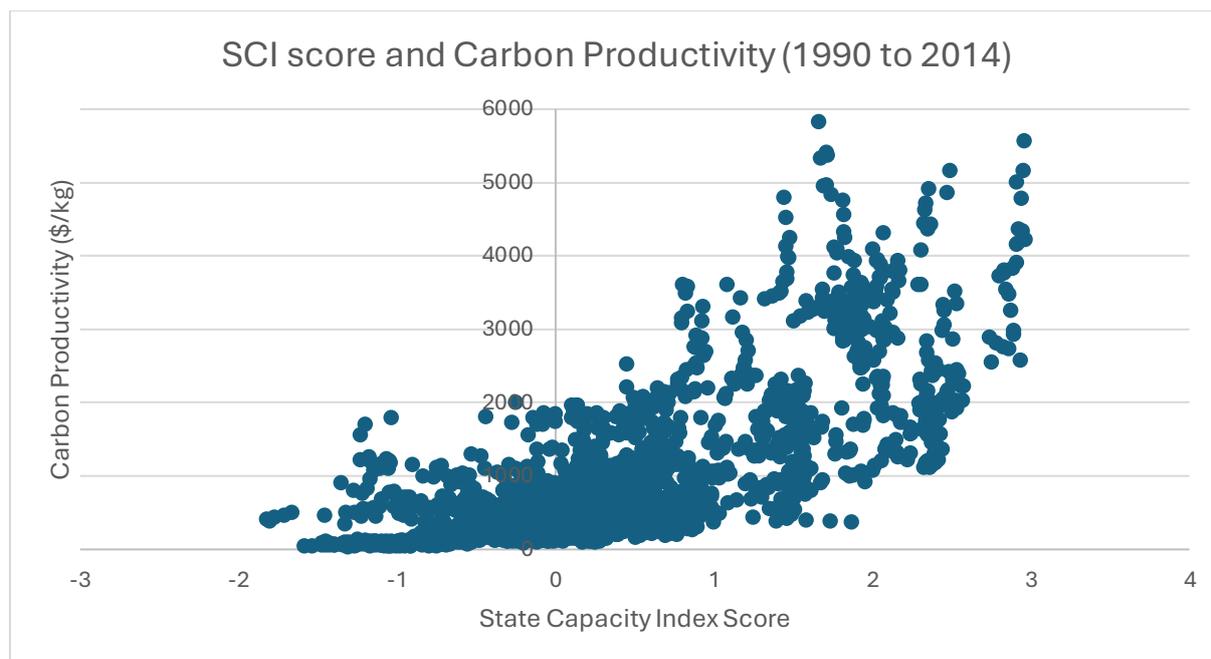

*Chart 1.2 : SCI score and Carbon Productivity*

These visuals suggests that developed countries (i.e. those with higher HDI/SCI) tend to have a less consistent correlation between state of development and Carbon Productivity. Further, we may infer carbon productivity as a universal exponential function of development that decays for large development (also see *Charts 1.3* and *1.4*); consistent with our initial goals, we have chosen to proceed with a two-period framework to assess the strength of correlation between development and Carbon Productivity.

For the purpose of this research, we consider the performance of G7 member countries to be representative of the 'developed' period, and an arbitrary selection of highly visible 'developing' countries (described in Appendix I) to stand in for the alternative period.

1. **Inference from the model for G7 observations:**

    There appears to be a very weak relationship between Carbon productivity and HDI in developed countries, as underscored by a low R-squared value of 0.370. This may be attributed to the fact that developed countries have smaller gains to extract from improved carbon productivity in improving their HDI. Exact statistics have been included in *Table 1.1*.

2. **Inferences for developing countries:**

    Here we find a more direct relationship between HDI and carbon productivity, as indicated by a higher R-squared value. These results are in line with our contention that increased education, standard of living, improved human capital, etc. (the determinants of HDI) represent a strong positive effect of improved carbon

productivity. This can be verified by the highly significant t-statistic under 'Developing Economies' in the *Table 1.1*.

We also note that this result holds when a control for state capacity has been accounted for in the regression model, thereby preventing the problem of omitted variable bias, or HDI accounting for the effect of state capacity in relation to outcomes for carbon productivity.

The statistically and economically significant coefficient of HDI (2,874.6) suggests a very strong causal relationship between HDI and carbon productivity and reinforces the argument for developed countries to provide climate finance for developing countries (an issue we discuss further in section 3).

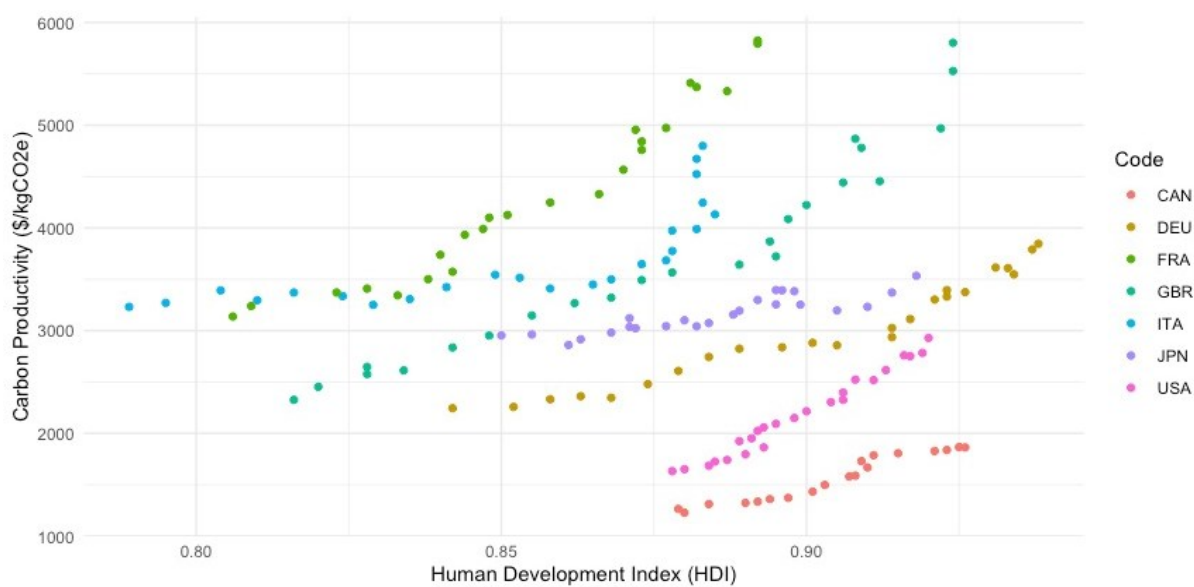

*Chart 1.3 : G7 Dataset*

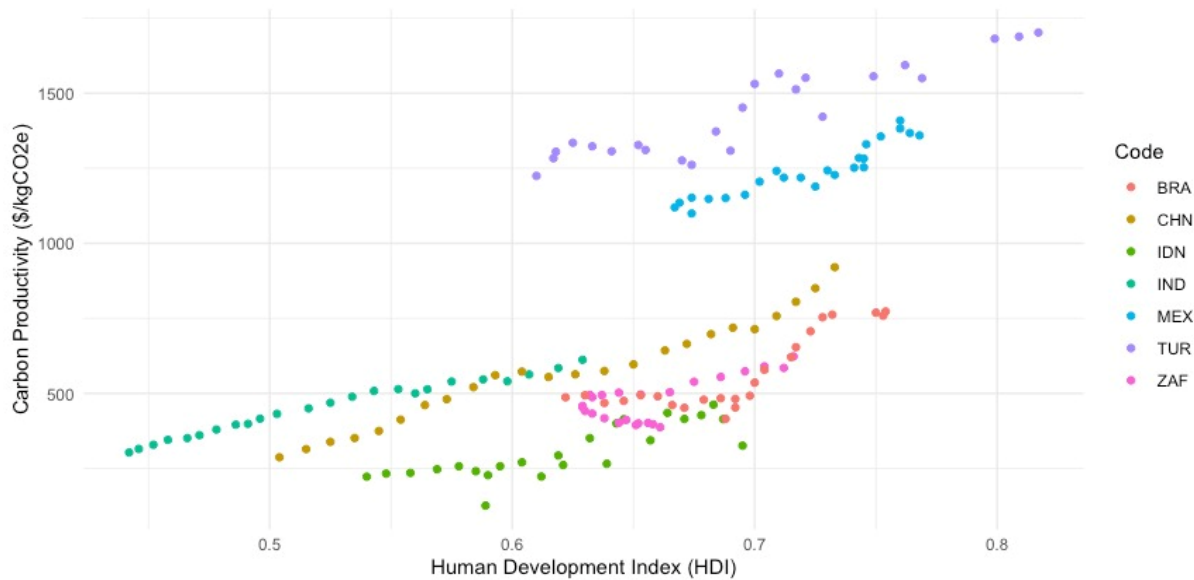

*Chart 1.4 : Developing Economies Dataset*

| Carbon Productivity (dependent, US$/kg$_{CO2e}$) | G7 | Developing Economies |
|---|---|---|
| HDI | 6,246.248** | 2,874.627*** |
| | (2,194.858) | (349.899) |
| Control | −3,405.347*** | 268.691** |
| | (352.708) | (83.610) |
| Constant | 4,001.001* | −1,347.447*** |
| | (1,793.530) | (205.141) |
| Observations | 163 | 168 |
| $R^2$ | 0.370 | 0.449 |
| Adjusted $R^2$ | 0.362 | 0.442 |
| Residual Std. Error | 829.148 | 313.317 |
| | (df = 160) | (df = 165) |
| F Statistic | 46.977*** | 67.237*** |
| | (df = 2, 160) | (df = 2, 165) |
| Note: | *p<0.05; **p<0.01; ***p<0.001 | |

Table 1.1 : Regression Statistics for comparing development and carbon productivity

## 2 – Exploring links between accumulated impact of climate investment and the effectiveness of additional investment: Overview

The two-period framework (introduced in section 1B) implies that developed economies operate at a different equilibrium to that of developing countries, as supported by observations in *Charts 1.3* and *1.4*. Further, the dichotomy between observations in developing economies and their developed counterparts bears striking resemblance to features of steady state in Solow's model of economic growth (Solow, 1956); where per capita capital and per capita output are tied to each other in the traditional interpretation of Solow's model, connecting per capita capital (investment) and carbon productivity in much the same manner supports the inference that development and carbon productivity cease to demonstrate significant correlation after a certain combination of carbon productivity and development (unique to the economy in question) is approached.

Building on Solow's ideas, we can consider that developing economies engage in capital widening when investing in carbon productivity, while developed economies focus on capital deepening; i.e. any extra investment in developing economies induces successive improvement in carbon technology with no real external impetus, while any improvement of carbon productivity in developed economies will require significant strategic planning and executive oversight. In more common parlance: the developing world has nowhere to go but up. We carry this train of thought further by considering the fact that the developed world experiences a significant advantage in terms of cutting-edge climate research and a strategic approach to mitigating the impact of climate change, despite developing economies facing the brunt of adverse climate phenomena; 'financial support for mitigation actions undertaken by developing countries has mainly been project-based' (Neuhoff, et al., 2009).

According to the Solow analogy, any and all climate investment in developing countries will have real and immediate grassroots impact, and the benefits of this impact will be reaped through development over and above that spurred by initial development; while this represents a net improvement for people living far below the global steady state, it leads to complacency and frustration in the developing world. While policymakers in developed economies may raise objection to the apparent cost of deliberate climate action, the developing world may be plagued by lack of political will due to an abundance of benefit from haphazard, arbitrary climate investment. Although complacence of this nature will serve a significant portion of the world's population just fine for the foreseeable future, it will also snowball as a free-rider problem where a critical mass of economies approaching a carbon productivity steady state may expect solutions from the 'next-in-line'; given sufficient free-riders, this critical mass could gradually spill over into geopolitical confrontation over resources and accountability, conflict, and even total stoppage of international trade.

In essence, climate investment does appear to demonstrate diminishing marginal utility when conducted as a standalone activity at scale; however, the diminishing marginal utility of climate investment can also be a signal for policymakers to develop long-term climate action plans to effect co-ordinated climate investment taking a holistic view of economic factors.

## 3 – Discussing the potential climate impact of additional investment in terms of the domestic state of economic development: Policy Critique

Having established the viability of a two-period framework for describing links between carbon productivity, an economy's state of development, and the scope of climate impact as a result of an additional unit of climate investment, it is now necessary to consider how such a framework would inform policy decisions. At the outset, our findings appear to suggest that the best way forward for developed economies is to allocate climate investment in a manner that finances action in developing countries; such an approach would yield the two-fold climate benefit of quick results and averting emissions that would otherwise be released as a consequence of development, due to underproductive carbon-intensive capital. In other words, any investment outside of the developing world comes with significant opportunity cost in terms of overall capital efficiency for development in the developing world, and therefore adverse climate impact on a global scale.

More nuanced inference suggests that significant channelling of investment from developed economies to their developing counterparts will only be viable for sporadic, discrete sets of time. The rationale for this is three-fold:

1. Potential for 'immediate' improvement (and attached opportunity cost of non-investment) will progressively diminish with investment until both considerations become insignificant relative to the developed world.

2. Global institutions and diplomatic forums in their current form are unprepared (and will likely be unable) to provide and sustain incentives that induce fair participation on the part of developed economies; investment as such would essentially entail developed economies handing competitive advantage to their developing counterparts on a silver platter.

3. Development is not necessarily a simple function of investment in domestic capabilities; throwing money at an economic problem is a recipe for disaster. Whether in the form of hyper-inflation, simple friction, or even some hitherto unobserved combination of systemic factors, economics will catch up with all inorganic investment and inevitably render further investment ineffective.

Economies with high carbon productivity can instead be treated as sinks for voluntary international investment, much as an innovative business would attract capital from financial markets; similarly, economies with significant catch-up potential can be allowed to focus on just that, while their developed counterparts lead innovation. By allowing developed countries to act as value-adding borrower-spenders, while developing economies play the role of lender-savers identifying suitable lending opportunities, a policy framework as such would encourage free and fair markets for information and investment to emerge organically.

Further, securities can be engineered to channel funds as a financial investment would in monetary terms, reframing the 'good' of climate investment in standard economic terms for better clarity of how money is spent. In addition to improving transparency, securitising climate investment will foster greater international cooperation, as economies will find themselves in a position to meet their climate commitments through the best investment possible, as opposed to choosing domestic prospects that may well be uneconomical given their domestic opportunity cost. Through democratisation of climate investment in this manner, developing countries that will generally be more affected by the consequences of climate change (Chinowsky, et al., 2011) will be empowered to invest in cutting-edge innovation by their developed counterparts that may otherwise be unviable for a catching-up economy to undertake; this will further improve allocative efficiency, as innovation addressing issues faced by a large section of the world will in fact receive a large portion of investment.

A securitised approach also addresses the free-lunch dilemma (discussed earlier in this section) from the developed world to developing countries, as developed economies can identify and finance the most optimal catch-up efforts offering far greater immediate return than domestic cutting-edge effort. Economies that would otherwise be classified as 'developing' may seek investment for cutting-edge research, which will be financed if relevant; similarly, economies in the 'developed' period are not bound by monetary obligations to their 'developing' counterparts unless the former seeks investment; financing channels lay the groundwork for implementing long-term cutting-edge innovation by providing means for catch-up development in the short term. Rigid classification is made redundant, and economies can choose which side of a transaction they are best suited to participate from.

At its core, international policy based on a two-period framework builds upon 3 key characteristics of Solow's ideas:

1. Distinct periods of equilibrium where one complements the other.
2. Dichotomous dynamic between equilibria.
3. Desirability of one period over the other [1].

Naturally, there also must be a goal that the international community seeks to achieve. Current consensus surrounds the need for climate action, and we have framed our research with climate investment in mind. It is interesting to consider how a two-period framework may apply to circumstance society may find itself facing in the decades to come; with space exploration rapidly becoming significant in the politics of national interest, we may soon face similar dilemmas to do with fair access to space and space exploration. The developed-developing differential has been a hallmark of human society since time immemorial; whether it manifests through economic haves and have-nots or otherwise, understanding this dichotomy and building upon the real transactions it yields will empower policymakers – for better, or for worse – to institute organic, dynamic frameworks that shape collaborative action at scale.

---

[1] Note the implications of desirability, as opposed to objective superiority.

## Appendix I

The following countries have been used to stand in for economies categorised as 'developing' in this study.

- Federative Republic of Brazil (Brazil)
- People's Republic of China (China)
- Republic of India (India)
- Republic of Indonesia (Indonesia)
- United Mexican States (Mexico)
- Republic of South Africa (South Africa)
- Republic of Türkiye (Türkiye)

We acknowledge that this list is not exhaustive, and may not be representative of the developing world in all contexts. This selection has been motivated by a number of subjective factors including visibility of data, consistent categorisation as 'developing', and relative stability of policy and administration.

## Appendix II

Table A.1: Ramsey's RESET test

| Model Dataset | RESET | df_1 | df_2 | p-value |
|---|---|---|---|---|
| G7 | 15.453 | 2 | 158 | 7.419e-07 |
| Developing Economies | 15.302 | 2 | 163 | 8.123e-07 |

Table 1.1: Regression statistics

| Carbon Productivity (dependent, US$/kg$_{CO_{2e}}$) | G7 | Developing Economies |
|---|---|---|
| HDI | 6,246.248** (2,194.858) | 2,874.627*** (349.899) |
| Control | −3,405.347*** (352.708) | 268.691** (83.610) |
| Constant | 4,001.001* (1,793.530) | −1,347.447*** (205.141) |
| Observations | 163 | 168 |
| $R^2$ | 0.370 | 0.449 |
| Adjusted $R^2$ | 0.362 | 0.442 |
| Residual Std. Error | 829.148 (df = 160) | 313.317 (df = 165) |
| F Statistic | 46.977*** (df = 2, 160) | 67.237*** (df = 2; 165) |
| Note: | *p<0.05; **p<0.01; ***p<0.001 | |

## Index of Quoted Figures

Listed in order of first appearance: